\documentclass[12pt]{article}

\usepackage{epsfig}

\usepackage{amssymb}
\usepackage{amsmath}
\usepackage{amsfonts}

\usepackage{amsthm}

\theoremstyle{theorem}

\theoremstyle{definition}

\def\bp{\begin{proof}}
\def\ep{\end{proof}}

\def\be{\begin{equation}}
\def\ee{\end{equation}}
\def\ba{\begin{array}{c}}
\def\ea{\end{array}}

\def\ben{$$}
\def\een{$$}

\newcommand{\bea}{\begin{eqnarray}}
\newcommand{\eea}{\end{eqnarray}}

\newcommand{\kt}{\rangle}

\begin{document}

\titlepage

\vspace{.35cm}

 \begin{center}{\Large \bf

Quantum star-graph analogues

of ${\cal PT}-$symmetric square wells

  }\end{center}

\vspace{10mm}

 \begin{center}

 {\bf Miloslav Znojil}

 \vspace{3mm}
Nuclear Physics Institute ASCR,

250 68 \v{R}e\v{z}, Czech Republic

{e-mail: znojil@ujf.cas.cz}

\vspace{3mm}

\end{center}

\vspace{5mm}

\section*{Abstract}

Non-Hermitian ${\cal PT}-$symmetric Hamiltonians $H=-d^2/dx^2+V(x)$
with $x \in \mathbb{R}$ are reinterpreted as describing the most
elementary phenomenological quantum graph, i.e., a system living on
the two half-line edges connected at a single matching-point vertex
in the origin. A $q-$pointed star graph generalization of these
$q=2$ models is then proposed and studied. For a special toy-model
point interaction yielding the exactly solvable model at $q=2$, the
bound-state energies are finally identified with the roots of a
remarkably compact trigonometric function at any $q=2,3,\ldots$.

\section*{PACS}

03.65.Ca Formalism

03.65.Db Functional analytical methods

03.65.Ta Foundations of quantum mechanics; measurement theory

03.70.+k Theory of quantized fields

\newpage


\section{Introduction}

The heuristic use of the concept of ${\cal PT}$ symmetry, i.e., of
the parity-times-time-reversal symmetry of quantum Hamiltonians $H$
and/or of the toy-model wave functions $\psi(x)$ (with, say, $x \in
\mathbb{R}$) proved unexpectedly productive in phenomenologically
oriented quantum field theory \cite{BM,Carl} or in the context of
relativistic quantum mechanics \cite{sm,ali}, in the supersymmetric
model-building \cite{ptsusy,pts} or, recently, in experimental
classical optics \cite{Makris}.

One of the simplest illustrative examples of a ${\cal PT}$ symmetric
Hamiltonian has been proposed in Ref.~\cite{david}. In the model the
quantum motion remained free inside a finite interval of
coordinates,
 \be
 -\frac{d^2}{dx_{}^2}\Psi_{}(x_{})
 = E\,\Psi_{}(x_{})
 \,,\ \ \ \ \ \ \
 x_{}\in (-L,L)\,.
 \label{duoplich}
 \ee
The only dynamical information has been carried by the very specific
point interaction induced by the external Robin-type boundary
conditions containing the single real coupling constant $\alpha$,
 \begin{equation}\label{bc}
  \Psi'(\pm L) + i \alpha \;\! \Psi(\pm L) = 0\,,
  \ \ \ \ \alpha>0
  \,.
 \end{equation}
This dynamical input yielded the  phenomenology-oriented real
bound-state spectrum
\begin{equation}
\label{spectrum}
 E_0=\alpha^2\,,
 \ \ \
 E_n=\left (\frac{n\pi}{2L} \right )^2\,,
 \ \ \
 n=1,2,\ldots \,.
\end{equation}
Its explicit form enabled us to restrict our attention, for the sake
of simplicity, to the non-degenerate systems where $2L\alpha/\pi
\neq 1,2,\ldots$. The probabilistic quantum-mechanical
interpretation of the closed-form wave functions of the model also
appeared feasible. The explicit formulae yielding the unitary forms
of the model (i.e., in the notation of review \cite{SIGMA}, {\em
all} of the non-equivalent ``standard'' inner-product
representations of the eligible physical Hilbert space of states
${\cal H}^{(S)}$) were found and described in a series of
mathematically rigorous subsequent studies \cite{Siegl,siegldis}.

The appealing and, in some sense, extreme simplicity of the latter
exceptional model inspired, naturally, a number of generalizations
(cf., e.g., Refs.~\cite{siegldis,Sieglbe}). Also our present paper
will describe a new generalization of the model.

\section{A quantum-graph reinterpretation of the ${\cal PT}-$symmetric
 square-well models}

A {\em formal} core of our present considerations will lie in the
reinterpretation of Eq.~(\ref{duoplich}) where the interval of
$x_{}\in (-L,L)$ will be treated as a union $\mathbb{G}^{(2)}$ of a
pair of equal-length subintervals (or ``edges'') $e_+= (0,L)$ and
$e_-= (-L,0)$ forming an elementary ``graph'' with the single
``vertex'' at $x=0$.

In such a case it is necessary to distinguish between the
theoretical and purely phenomenological {\em informal} aspects of
such a reinterpretation. Indeed, the latter, ``realistic'' aspect is
very natural. Traditionally, it finds its widespread use in quantum
chemistry where, typically, the valence electron of an organic
molecule may be often treated as moving just strictly along the
atomic-bond edges \cite{[4]}.

In the former, more abstract and less phenomenological setting the
restriction of the motion to the edges of a suitable graph may
enormously simplify the underlying Schr\"{o}dinger equation
\cite{[7]}. Recently, this idea made the study of various
quantum-graph models extremely popular. {\it Pars pro toto} the
interested reader may be recommended to consult a comprehensive
collection \cite{[6]} of more than 700 pages of reviews and original
research reports, with the scope ranging from certain entirely
``unrealistic'' scenarios (i.e., e.g., from the fractal and/or
chaos-simulating graphs \cite{[12]}) down to certain very realistic
models of observable photonic crystals and various other
``leaky-graph'' nanostructures encountered, typically, in condensed
matter physics \cite{[8]}.

For the sake of definiteness, let us now assume that in our above
most elementary graph $\mathbb{G}^{(2)}$, both of the respective
edges are oriented {\em inwards}, i.e., $e_\pm =e_\pm (y_\pm )$ with
$y_+=L-x \in (0,L)$ while $y_-=L+x\in (0,L)$. Without any real loss
of generality, our attention will also remain restricted to the
dynamics represented by the end-point point interaction as mediated
by boundary conditions (\ref{bc}).

In the new notation we have to replace Eq.~(\ref{duoplich}) by the
pair of differential Schr\"{o}dinger equations
 \be
 -\frac{d^2}{dy_{\pm}^2}\psi_{\pm}(y_{\pm})
 = E\,\psi_{\pm}(y_{\pm})
 \,,\ \ \ \ \ \ \
 y_{\pm}\in (0,L)\,
 \label{dvaplich}
 \ee
complemented by the standard regular matching conditions in the
origin,
 \begin{equation}
 \label{bcin}
  \psi_{+}(L)=\psi_{-}(L)\,,
  \ \ \ \ \ \
  \partial_{y_{+}}\psi_{+}(L)+\partial_{y_{-}}\psi_{-}(L)=0\,.
  \,
 \end{equation}
The external Robin-type boundary conditions (\ref{bc}) must now
read, {\it mutatis mutandis},
 \begin{equation}\label{bcr}
  \psi'_{\pm}(0)= \pm i \alpha \;\! \psi_{\pm}(0)\,.
 \end{equation}
The physics (i.e., the spectrum) remains unchanged but the
mathematical meaning of the ${\cal PT}-$symmetry of $H$ (i.e., the
representation of the antilinear operator $\omega={\cal PT}$) gets
modified.

\subsection{Operators of symmetries}

After the change of the language, the differential-operator
Hamiltonian (originally defined as acting, in general, in
$L^2(\mathbb{R})$) must be treated as acting in another,
``friendly'' \cite{SIGMA} Hilbert space of states ${\cal
H}^{(F)}=L^2(\mathbb{R}^+)\bigoplus L^2(\mathbb{R}^+)$. Formally,
Schr\"{o}dinger Eq.~(\ref{dvaplich}) then acquires the two-by-two
operator-matrix form
 \be
 \left(
 \begin{array}{cc}
 H_+-E&0\\
 0&H_--E
 \ea
 \right)
 \left (
 \ba
 \psi_+\\
 \psi_-
 \ea
 \right )=0\,
 \ee
(i.e., $(H-EI)|\psi\kt=0$ in an abbreviated notation).

The original {\em linear} operator of parity ${\cal P}$ (i.e., the
reflection which changed the sign of the coordinate, ${\cal P}: x
\to -x$, $x \in (-L,L)$) will now play the slightly different role
of a domain-intertwiner such that ${\cal P}: y_\pm \to y_\mp$, i.e.,
 \be
 {\cal P}=
 \left(
 \begin{array}{cc}
 0&I\\
 I&0
 \ea
 \right)
 \,.
 \ee
The parallel interpretation of the {\em antilinear} symmetry
$\omega={\cal PT}$ of $H$ will vary with the spectral properties of
$H$ \cite{Wigner}. As long as we may define $|\psi'\kt =\omega
|\psi\kt$, we may write $\omega(H-EI)|\psi\kt=(H-E^*I) |\psi'\kt
=0$. Thus, in the generic non-degenerate case we must distinguish
between the real-energy scenario $E=E^*$ (in which $|\psi'\kt$ will
be proportional to $|\psi\kt$) and the case of $E\neq E^*$ in which
the two eigenvectors $|\psi'\kt$ and $|\psi\kt$ of $H$ remain
linearly independent (or, in the language of wave functions, in
which the ${\cal PT}$ symmetry becomes spontaneously broken
\cite{Carl}).

\subsection{Wave functions and energies}

In the case of the unbroken ${\cal PT}$ symmetry we may always
change the phase of the initial ket vector $|\psi\kt$ in such a way
that $|\psi'\kt=|\psi\kt$. In the original $L^2(\mathbb{R})$ context
this was a normalization convention in which both the respective
symmetric and antisymmetric components of $\Psi(x) = S(x) + {\rm
i}A(x)$ with properties $S(-x)=S(x)$ and $A(-x)=-A(x)$ were real.

In the current literature, people sometimes speak about the ${\cal
PT}$ symmetry of the system while {\em tacitly assuming} that it is
{\em not} spontaneously broken, i.e., that the spectrum is real and
non-degenerate. Under such an assumption it is rather
straightforward to return to our specific model and to write down
the definitions of the two new wave functions $\psi_\pm(y_\pm)$ with
$y_\pm \in (0,L)$ in terms of the components of the old wave
function $\Psi(x)$,
 \be
 \psi_-(y_-)=S(L-y_-)-{\rm i}A(L-y_-)\,,
 \ \ \ \ \ \
 \psi_+(y_+)=S(L-y_+)+{\rm i}A(L-y_+)\,.
 \ee
This formula confirms that the time reversal operator ${\cal T}$
itself acts, as usual, as complex conjugation.

In the new notation the general solution of differential
Eq.~(\ref{dvaplich})
 \be
 \psi_\pm(y)=A_\pm \,\sin ky + B_\pm\,\cos ky\,
 \label{val2}
 \ee
must be restricted, first of all, by the external boundary
conditions at $y=0$. This yields the rule
 \be
 k\,A_\pm = \pm {\rm i}\alpha B_\pm \,.
 \ee
Its insertion in Eq.~(\ref{val2}) defines the wave functions.
Finally, the necessity of their matching in the central vertex,
i.e., relation
 \be
  \label{bgein2}
  \frac{
 k\,\tan kL-{\rm i}\alpha}{k+ {\rm i}\, \alpha \tan kL
    }+ \frac{
 k\,\tan kL+{\rm i}\alpha}{k- {\rm i}\, \alpha \tan kL
    }=0\,
  \,
 \ee
leads to the ultimate secular equation
 \be
  \label{sece2}
 2\,\,
 \frac{k^2-\alpha^2}{k^2+\alpha^2\,\tan^2 kL}
 \tan kL=0\,.
 \ee
Although this equation looks different from the secular equation as
given in Ref.~\cite{david} (where one merely has to set $d=2\,L$ in
eqs. Nr. 13 and 14), the set of the resulting energy roots
(\ref{spectrum}) remains the same of course. It is worth noticing
that from our present form of secular equation~(\ref{sece2}) the
complete set of eigenvalues is determined via zeros of a triplet of
elementary functions $f_1(k):=k^2-\alpha^2$, $f_2(k):=\tan kL$  and
$f_3(k):= {\rm cotan}\, kL $.

\section{The new model with $q$ equilateral edges}

In the above-introduced notation the generalization of the model
becomes straightforward. At any $q\geq 2$ we merely consider a
$q-$plet of Schr\"{o}dinger equations
 \be
 -\frac{d^2}{dy_{j}^2}\psi_{j}(y_{j})
 = E\,\psi_{j}(y_{j})
 \,,\ \ \ \ \ \ \
 y_{j}\in (0,L)\,,
 \ \ \ \ \ j = 0, 1, \ldots, q-1
 \label{kveplich}
 \ee
for which the $q-$plet of edges $e_j =e_j (y_j )$ with $y_j\in
(0,L)$ may be visualized as forming a star-shaped graph
$\mathbb{G}^{(q)}$ with the single central vertex at $y_j=L$. For
the sake of simplicity, the matching in the origin will be chosen in
the elementary Kirchhoff's form
 \begin{equation}
 \label{bcinge}
  \psi_{j}(L)=\psi_{0}(L)\,,\ \ \ \ j =  1,2, \ldots, q-1\,,
  \ \ \ \ \ \ \
  \sum_{j=0}^{q-1}\,\partial_{y_{j}}\psi_{j}(L)=0
  \,.
 \end{equation}
In a completion of the tentative analogy, the complex rotation by
angle $\pi$ as used in Eq.~(\ref{bcr}) will be replaced now by the
complex rotation by an appropriate fractional angle
$\phi=\phi(q)=2\pi/q$, yielding the prescription
 \begin{equation}
 \label{brody}
  \partial_{x_{j}}\psi_{j}(0)= i \alpha\, e^{{\rm i}j\varphi}\,
  \psi_{j}(0)\,,
  \ \ \ \ \ j = 0, 1, \ldots, q-1\,,
  \ \ \ \varphi=2\pi/q
  \,.
 \end{equation}

\subsection{Wave functions}

The general solution of the differential Schr\"{o}dinger system
(\ref{kveplich})
 \be
 \psi_j(x)=A_j \,\sin kx + B_j\,\cos kx
 \,,\ \ \ \ j = 0, 1, \ldots, q-1
 \label{val}
 \ee
yields also the auxiliary expression for the derivatives,
 \be
 \partial_x\,\psi_j(x)=k\,A_j \,\cos kx - k\,B_j\,\sin kx
 \,,\ \ \ \ j = 0, 1, \ldots, q-1\,.
 \label{valpr}
 \ee
One converts the dynamical boundary conditions (\ref{brody}) into an
elementary connection between coefficients,
 \begin{equation}
 \label{brodyb}
 k\,A_j
  = i \alpha\, e^{{\rm i}j\varphi}\,B_j
  \,,
  \ \ \ \ \ j = 0, 1, \ldots, q-1\,,
  \ \ \ \varphi=2\pi/q
  \,.
 \end{equation}
The continuity condition for wave functions in the central vertex
 \begin{equation}
 \label{bcingeljf}
  \left [ i \alpha\, e^{{\rm i}j\varphi}
   \,\sin kL + k\,\cos kL\right ]\,B_j=k\,\varrho\,,\ \ \ \
   j = 0, 1, \ldots, q-1\,
 \end{equation}
enables us to define all of the coefficients $B_j=B_j(\varrho,L,k)$
as proportional to the auxiliary parameter $\varrho$. Their
subsequent insertion in the explicit version
 \be
  \label{bgeljf}
   \sum_{j=0}^{q-1}\,\left [ i \alpha\, e^{{\rm i}j\varphi} \,\cos kL -
 k\,\sin kL\,\right ]\,B_j=0
  \,
 \ee
of the Kirchhoff's law of Eq.~(\ref{bcinge}) finally leads to a
rather complicated trigonometric secular equation which defines, in
principle at least, all of the bound-state energies $E=E_n$. As long
as the underlying Hamiltonian is non-Hermitian, these energies may
be both real and complex at $q>2$. Some of them also need not remain
expressible via any closed-form analogue of the special $q=2$
formula~(\ref{spectrum}).

\subsection{Secular equation}

After the elimination of $B_j$s the secular equation for bound-state
energies $E=k^2$ acquires a compactified form
 \be
  \label{bgeljfin}
  \sum_{j=0}^{q-1}\,
 \tan \left [kL- \beta_j(k)\right ]=0\,
  \,
 \ee
where the parameter $\varrho$ dropped out and where we defined,
implicitly,
 \ben
   \tan \beta_j(k) = \frac{{\rm i} \alpha\, \exp({{\rm
   i}j\varphi})}{k}=C+{\rm i}K
   \,,\ \ \ \ j = 0, 1, \ldots, q-1\,,
  \ \ \ \varphi=2\pi/q\,.
 \een
Setting $\beta_j=u+{\rm i}v$ we may abbreviate $\tan u=U$ and $\tanh
v =V $ and write
 \ben
 \tan \beta_j=\frac{U+{\rm i}V}{1-{\rm i}UV}=
 \frac{(U+{\rm i}V)(1+{\rm i}UV)}{1+U^2V^2}=
 \frac{U(1-V^2)+{\rm i}V(1+U^2)}{1+U^2V^2}
 \,.
 \een
As long as
%
 \ben
 C=C(j,k)=-\frac{ \alpha}{k}
 \,\sin j\varphi\,,
 \ \ \ \ \ \
 K=K(j,k)=\frac{ \alpha}{k}
 \cos j\varphi\,
 \een
%
%
we obtain the real-function correspondences
 \ben
 C(j,k)=\frac{\sin u \cos u}{\cos ^2 u + \sinh^2 v}\,,
 \ \ \ \
 K(j,k)=\frac{\sinh v \cosh v}{\sinh^2 v+\cos ^2 u }\,
 \een
as well as the closed-form inversion formulae
 \ben
  \tan j\varphi
 =\frac{\sin 2u}{\sinh 2v}\ \left (\,=\frac{C}{K} \right )\,,
 \ \ \ \ \
 \frac{\alpha^2}{k^2}
 =\frac{\sinh^2 v + \sin^2 u}{\sinh^2 v + \cos^2 u}
 \  \left (\,=C^2+K^2 \right )\,.
 \een
From the latter relation we may finally eliminate  $\sinh v$ and
convert the former relation into an easily solvable quadratic
equation for the value of $\cos^2 u$, yielding the two eligible
roots as functions of $j$ and $k$.

In this manner, secular equation (\ref{bgeljfin}) would be given a
lengthy and rather clumsy but still explicit elementary form which
we are not going to display here of course. Anyhow, at any pair of
given parameters $q$ and $\alpha$ the whole eigenvalue problem may
be now solved, numerically, with arbitrary precision.



\section{Secular equation revisited\label{simpler}}

Let us now demonstrate that for the purposes of
symbolic-manipulation simplifications, secular Eq.~(\ref{bgeljfin})
should be reconsidered in the apparently more complicated form of
the sum
 \be
  \label{bgein}
   \sum_{j=0}^{q-1}\,\frac{
 \tan kL-C(j,k)-{\rm i}K(j,k)}{1+ [C(j,k)+{\rm i}K(j,k)]
   \,\tan kL }=0\,.
  \,
 \ee
In what follows we are now going to show that and how the
simplification of this formula may be achieved by the explicit
summation when proceeding, in a systematic inductive manner, from
the smallest integers $q$ upwards.

\subsection{The trivial single-line quantum graph: $q= 2$. \label{tento}}

At $q=2$ we have $\tan \beta_{0,1}(k)=\pm {\rm i}\alpha/k$ so that
one obtains $C(j,k)=0$ and $K(\{0,1\},k)=\pm \alpha/k$. This makes
secular equation (\ref{bgein}) reducible to the ultimate elementary
constraint (\ref{sece2}).

\subsection{The three-pointed star graph}

When we abbreviate $\gamma=\alpha/2$, the $q=3$ three-term secular
equation
 \be
  \label{bgein3}
  \frac{
 k\,\tan kL-2{\rm i}\gamma}{k+ 2{\rm i}\, \gamma \tan kL
    }+ \frac{
 k\,\tan kL +\gamma\,\sqrt{3}+{\rm i}\gamma}{k- ({\rm i}+\sqrt{3})\,
  \gamma \tan kL
    }+ \frac{
 k\,\tan kL -\gamma\,\sqrt{3}+{\rm i}\gamma }{k- ({\rm i}-\sqrt{3})\,
  \gamma \tan kL
    }
    =0\,
  \,
 \ee
may be rewritten in its simplified, single-term form
 \be
  \label{bgein3s}
  3\,\frac{
 k^3+{\rm i}\alpha^3\, \tan kL}{k^3-{\rm i}\, \alpha^3 \tan^3 kL
    }\,\tan kL
    =0\,.
  \,
 \ee
The inspection of this secular equation reveals that the real and
discrete part of the $q=3$ bound-state spectrum coincides with its
$q=2$ predecessor, up to the anomalous root $k=\alpha$ which now
disappeared. In other words, the real roots of the new secular
Eq.~(\ref{bgein3s}) coincide now strictly with the zeros of the real
functions $\tan kL$ and ${\rm cotan}\, kL$.

It is necessary to add that our secular Eq.~(\ref{bgein3s}) also
possesses complex roots defined by subcondition
 \be
  \label{com3s}
  k^3+{\rm i}\alpha^3\, \tan kL
    =0\,.
  \,
 \ee
We may decompose ${k}/{\alpha}=\mu+{\rm i}\nu$, set
$\lambda=\alpha\,L$ and obtain the complex version of such a secular
subequation
 \ben
 \mu^3 - 3\,\mu\,\nu^2
 +{\rm i}
 \left (3\,\mu^2\nu-\nu^3
 \right )
 =
 \frac{\tan \mu\lambda+{\rm i}\,\tanh \nu \lambda }
 {1-{\rm i}\tan \mu\lambda\,\tanh \nu \lambda}
 \een
which is equivalent to the coupled pair of the real secular
subequations
 \ben
 \mu^3 - 3\,\mu\,\nu^2+
 \left (3\,\mu^2\nu-\nu^3-1/\tanh \nu\lambda
 \right )\,\tan \mu\lambda\,\tanh \nu \lambda
 =
 0
 \een
and
 \ben
 3\,\mu^2\nu-\nu^3-\tanh \nu \lambda
  -\left (\mu^3 - 3\,\mu\,\nu^2\right )
  \,\tan \mu\lambda\,\tanh \nu \lambda
 =0
 \,.
 \een
In an extensive numerical test we revealed and demonstrated the
existence  of nontrivial complex roots of these equations at the
various values of $\lambda$. For example, we localized the sample
pair of roots with $\mu=1.20484$ and $\nu=\pm 0.3507$ at
$\lambda=1$. This means that at $q=3$ the non-Hermiticity of the
Hamiltonian may probably be interpreted as ``too strong''. In other
words, the Hamiltonian of the quantum version of this system cannot
be Hermitized using just the standard techniques as reviewed in
Ref.~\cite{SIGMA}, at any coupling constant $\lambda$.

In this sense, the practical phenomenological applicability of our
$q=3$ quantum graph appears restricted to the domain of non-linear
optics \cite{Makris} and to the various similar, recently popular
classical-physics (or even classical-mechanics \cite{wax})
implementations and applications of the theory where the complex
energies may and do find their natural physical interpretation.

Naturally, the situation is different in quantum physics where, {\em
inside the physical Hilbert space } ${\cal H}$, the spectrum of {\em
any} operator $H$ representing an observable quantity {\em must} be
real. Still, there exists a certain recently discovered~\cite{unbou}
space-projection trick which may prove acceptable in at least some
phenomenological considerations and applications of our models.

\subsection{Quantum-Hilbert-space construction at $q=3$}

We just demonstrated, constructively, that the discrete energy
spectrum of at least some of our present $q>2$ quantum-graph models
need not necessarily be all real. In {\em all} of these
``non-real-spectrum'' cases it seems necessary to discard the
underlying Hamiltonian $H=H^{(q)}(\alpha)$ as leading to non-unitary
evolution of the quantum system in question. At the same time, many
of the formal (e.g., solvability) as well as phenomenological (e.g.,
scattering-related \cite{Coron}) features of these and similar
models might seem appealing enough. For this reason, let us now
describe, briefly, one of the recently discovered and more or less
universal remedies of the apparent complex-energy shortcoming.

Firstly, let us remind the readers that within the rigorous quantum
theories one must often exclude {\em all} of the non-Hermitian
Hamiltonians $H$ (leading to the real spectrum or not) which cannot
be assigned a suitable Hilbert space in which they may be
reinterpreted (typically, via a suitable inner product
\cite{Carl,SIGMA}) as self-adjoint. In particular, this year it has
been proved \cite{Krejcirik} that in this manner it would be even
necessary to discard the popular imaginary cubic oscillator and many
other standard benchmark ${\cal PT}-$symmetric quantum models with
real spectra.

Naturally, this conclusion may seem rather surprising.
Unfortunately, it is based on the rigorous functional analysis and,
hence, mathematically valid. The essence of the apparent paradox has
been found in the inconsistency between the {\em a priori} choice of
the {\em same} domain ${\cal D}$ of $H$ {\em before} and {\em after}
the Hermitization,  ${\cal D}^{\rm (before)}(H)={\cal D}^{\rm
(after)}(H)$.

This conclusion may be perceived as indication of the way out of the
trap. Indeed, a physics-oriented and pragmatic (thought still
mathematically rigorous) way out of such a form of crisis of the
theory has been found, almost in parallel, in Ref.~\cite{unbou}. In
a simplified explanation it has been merely admitted that ${\cal
D}^{\rm (before)}(H) \neq {\cal D}^{\rm (after)}(H)$.

The resulting flexibility of the ``projection'' on the meaningful
vector space ${\cal D}^{\rm (after)}(H)$ of the correct physical
states enables us to construct the latter space simply as spanned by
{\em any} subset of the eigenvectors of the Hamiltonian $H$ in
question. In other words, the formal recipe as presented in
Ref.~\cite{unbou} may be simply read as just another version of the
innovative implementation of the abstract principles of quantum
theory where the correct Hilbert space ${\cal H}$ is determined {\em
dynamically} (plus, in the present case, in the mere real-energy
subspace).

Once we return now to our present specific $q=3$ quantum-graph
Hamiltonian $H$ where we choose, {\em a priori}, the usual and
friendly (but, in general, unphysical) Hilbert-space domain ${\cal
D}^{\rm (before)}(H)=\bigoplus_{j=0}^2\,L^2(e_j)$ (in this space, $H
\neq H^\dagger$ of course), we may now use just the slightly adapted
recipe of Ref.~\cite{unbou}. Thus, in essence, we have to construct
the correct vector space ${\cal D}^{\rm (after)}(H)$ (as well as its
bra-vector dual) as spanned just by the real-eigenvalue eigenvectors
of $H$ (or of $H^\dagger$, respectively).

We omit the further details here, summarizing that due to the
infinite number of the real eigenvalues at our disposal, the
dimension of ${\cal D}^{\rm (after)}(H)$ will remain infinite. The
immanent projector-operator nature of the whole construction may,
indeed, be perceived as far from trivial, with details lying,
certainly,  far beyond of the scope of our present paper. Thus, we
may only add that although, in the projected-space approach, the
subsequent Hermitization of the model remains entirely routine
\cite{SIGMA}, the physical interpretation or our star-shaped-graph
models remains the same as in the most elementary $q=2$ special
case. For this reason, some of the apparent paradoxes (like, e.g.,
the intrinsically non-local nature of the well known $q=2$ system
\cite{Coron}) will survive the transition to $q>2$ of course.

\subsection{Secular equation at $q=4$}

The  $q=4$ version of our secular equation reads
 \be
  \label{bgein4}
 \frac{
 k\,\tan kL-{\rm i}\alpha}{k+ {\rm i}\, \alpha \tan kL
    }+ \frac{
 k\,\tan kL+{\rm i}\alpha}{k- {\rm i}\, \alpha \tan kL
    }
    +
     \frac{
 k\,\tan kL-\alpha}{k+  \alpha \tan kL
    }+ \frac{
 k\,\tan kL+\alpha}{k- \, \alpha \tan kL
    }
    =0\,
  \,
 \ee
and may be again simplified,
 \be
  \label{bgein4s}
  4\,\frac{
 k^4+\alpha^4\, \tan^2 kL}{k^4- \alpha^4 \tan^4 kL
    }\,\tan kL
    =0\,.
  \,
 \ee
The real stable-bound-state spectrum remains the same as at $q=3$.
The complex roots of the auxiliary subequation $k^4+\alpha^4\,
\tan^2 kL=0$, i.e., of the two equations $k^2=\pm {\rm i} \alpha^2\,
\tan kL$ related by the formal change of $k \to -k$ may be sought
just in a half-plane of complex $k$. The final analysis of this
equation may again proceed in the manner outlined in preceding
subsection.

\begin{figure}[h]                     
\begin{center}                         
\epsfig{file=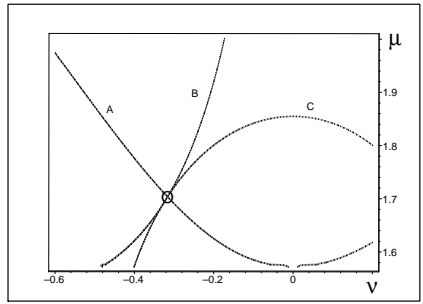,angle=270,width=0.4\textwidth}
\end{center}                         
\vspace{-2mm} \caption{Numerical identification of a complex root of
secular equation.
 \label{fedu}}
\end{figure}

Once we make a particular choice of the sign (say, plus), we obtain
the simplest special case of the equation for the complex roots
$k=\mu+{\rm i}\,\nu$. In units $L=1$ we have
 \be
 \mu^2-\nu^2+2\,{\rm i}\,\mu\nu= +{\rm i}\,\alpha^2\,
 \tan (\mu+{\rm i}\,\nu)
  \,.
 \label{orif}
 \ee
This is a complex equation which decays into the pair of real
conditions
 \be
 \mu^2-\nu^2+2\,\mu\nu\,\tan \mu\,\tanh \nu
 +\alpha^2\,
 \tanh \nu=0\,,
 \label{orifa}
 \ee
 \be
 2\,\mu\nu-(\mu^2-\nu^2)\,\tan \mu\,\tanh \nu
 -\alpha^2\,
 \tan \mu=0
  \,.
 \label{orifbe}
 \ee
The search for a nontrivial root of these two equations remains
numerical. Figure \ref{fedu} samples the localization of a {\em
complex} root $k=\sqrt{E}$ with components $\mu\approx 1.7025$ and
$\nu\approx -0.3165$ at $\alpha=1$.

Let us add that the efficiency as well as the reliability of the
latter search for a complex root was significantly enhanced by a
trick in which the two real equations (\ref{orifa}) and
(\ref{orifbe}) (defining, in our illustrative picture, curves A and
B in the $\mu-\nu$ plane, respectively) were complemented by the
third, redundant rule of the coincidence of the absolute values of
the left- and right-hand side of the complex relation $k^2=\pm {\rm
i} \alpha^2\, \tan k$,
 \be
 \mu^2+\nu^2=\alpha^2\,\sqrt{\frac {  \tan^{2}  \mu
  +  \tanh^{2}  \nu }{1+ \tan^{2}
 \mu
  \,\tanh^{2}  \nu }}\,.
 \label{orifcet}
 \ee
In our picture this defines the redundant, additional curve $C$ in
the $\mu-\nu$ plane. We see that its construction helps us to
identify the root in question via a {\em triple} intersection of the
curves A, B and C. Naturally, such a recipe keeps the numerical
errors under a very reliable control.

\subsection{Secular equation at $q=5$}

At $q=5$ our secular equation  becomes perceivably more complicated
but the use of computerized symbolic manipulations and appropriate
trigonometric identities is still found to lead to the thoroughly
simplified prescription
 \be
  \label{bgein5s}
  5\,\frac{
 k^5-{\rm i}\alpha^5\, \tan^3 kL}{k^5+{\rm i} \alpha^5 \tan^5 kL
    }\,\tan kL
    =0\,, \ \ \ \ \ q=5\,.
 \ee
This formula indicates that the real part of the spectrum remains
unchanged also at $q=5$. On the more important methodical level the
structure of this formula  confirms our expectation that the sum
(\ref{bgein}) may be represented by a very simple function of $q$.

\subsection{Secular equation at any $q$}

On the basis of the above particular results yielding the explicit
and elementary closed summation formulae it is straightforward to
conjecture and prove the validity of the following extrapolated
general trigonometric secular equation
 \be
  \label{bgeinqs}
  q\,\frac{
 k^q+({\rm i}\alpha)^q\, \tan^{q-2} kL}{k^q-({\rm i} \alpha)^q \tan^q kL
    }\,\tan kL
    =0\,, \ \ \ \ \ q= 2,3,\ldots\,.
 \ee
Whenever we restrict our attention just to the real roots $k$ which
correspond to the stable bound-state solutions, we reveal that the
numerator in the fractional part of the left-hand-side secular
determinant (\ref{bgeinqs}) plays now the role of the source of the
$q>2$ analogues of the single anomalous $q=2$ root $k=\alpha$.

This numerator cannot vanish at real $k$ and odd $q = 2m+1$, $m \in
\mathbb{Z}^+$ and  it cannot vanish at $q=4m$, $m \in \mathbb{Z}^+$,
either. The remaining values of the integer $q={4m-2}$, $m \in
\mathbb{Z}^+$ appear exceptional. Their choice leads to the
emergence of the additional real zeros and so it deserves a separate
attention.

\subsection{Secular equation at the exceptional $q=4m-2$}

At $q=2$ our quantum-graph spectrum was all real (cf. sec.
\ref{tento} above) but, as we saw, the situation became anomalous at
any $q\geq 3$. Nevertheless, what is new at the exceptional integers
$q=4m-2$, $m=2,3,\ldots$ is that the secular-equation factor
 \be
  \label{anonqsano}
   k^{4m-2}-\alpha^{4m-2}\, \tan^{{4m-4}} kL
    =0\,
 \ee
becomes nontrivial and, moreover, that it produces, obviously,
certain potentially real additional bound-state eigenvalues $E=k^2$.

One of the unfortunate consequences of the latter observation is
that some of the energies of the stable quantum-star bound states
{\em cease} to be obtainable in closed form. Their determination
must be performed by suitable brute-force numerical methods.
Moreover, the practical search for the roots of transcendental
Eq.~(\ref{anonqsano}) or of its slightly more friendly and
graphically better tractable version
 \be
  \label{anono}
   k_{anomalous}=\alpha\, \tan^{1-1/(2m-1)} k_{anomalous}  L
    =0\,, \ \ \ \ \ m= 1,2,\ldots\,
 \ee
becomes technically complicated. {\em A priori}, without any
extensive calculations we can immediately be sure that at the
sufficiently small $\alpha$s, all of the generic and $m-$independent
real roots $k=(n+1/2)\pi \gg 1$ with $n=0,1,\ldots$ become
accompanied by the neighboring real pairs of eigenvalues which are
produced by Eq.~(\ref{anono}).

With the growth of $\alpha$ this picture will first lose its
validity between $k=0$ and $k=\pi/2$. At the sufficiently small
$\alpha$s one always finds there the two smallest real roots inside
the interval. With the growth of parameter $\alpha$ these two roots
move towards each other at a speed which depends on $m$. In a
numerical experiment performed at $m=2$ we found that there exists
the critical value of $\alpha=\alpha_{critical} \approx 0.7863$ at
which these two lowest anomalous energy-level twins merge at $k
\approx 0.748$ and, subsequently, complexify.

This observation may only be read as a reliable numerical proof that
at the sufficiently large values of the strength of the
non-Hermiticity $\alpha> \alpha_{critical}$  (with
$\alpha_{critical} \approx 0.7863$ at $m=2$), the spectrum of the
whole system certainly contains non-real eigenvalues.

In the interval of $\alpha< 0.7863$ we may only conclude that there
exists a set of certain new and strictly real ``anomalous''
eigenvalues which may only be generated numerically (i.e., say, via
our secular sub-equation (\ref{anono})). This extremely interesting
infinite family of the new quantum states is, in principle,
observable. Its energy levels (which form, incidentally, almost
degenerate pairs at higher excitations) may be interpreted as the
appropriate quantum-graph $q=6$, $q=10$ (etc) analogues of their
single-state $q=2$ predecessor $E_0(\alpha)$ of
Eq.~(\ref{spectrum}).

%
%
%
%
%
%
%

\section{Summary}

In our preceding papers \cite{grII,int1} on ${\cal PT}-$symmetric
quantum graphs we always restricted our attention to their mere
discrete approximants. In our present paper, we abandoned this
approach as not sufficiently efficient. An alternative way of
circumventing the technical obstacles has been found here in an {\em
ad hoc} restriction of the class of the admissible graphs to their
star-shaped subset $\mathbb{G}^{(q)}$. Due to this restriction, we
were able to replace the universal though less powerful
discretization approach by the much more elementary method of
matching of wave functions at the central vertex.

In technical sense, our present results may be perceived as a return
to optimism. The main source of the simplification of our
constructive considerations may be identified with the inherent
symmetry of the complex Robin boundary conditions. This symmetry
found its fructification in the emergence of powerful trigonometric
identities. These identities led to the enormous simplification of
the related secular equations at any integer $q \geq 2$. In this
sense one could find here certain parallels with the role of
trigonometric identities, say, during the early stages of
development of Calogero models \cite{Calogero} and/or of some of
their less influential analogues \cite{Klee}.

The transition to nontrivial topology of the graph-related phase
space (or of the space of coordinates) manifested itself in two
ways. Firstly, a part of the physical sector where the bound-state
energies remained strictly real appeared {\em independent} of the
number of rays $q$, i.e., {\em mathematically} stable. Secondly,
strictly this part of the spectrum also remained defined by closed
formulae at $q>2$. In contrast, the rest of the spectrum (and, in
particular, the whole sector of ``resonances'' where the energies
are complex) appeared changing with the changes of $q$.

Due to the elementary form of the secular equations at any $q$, the
``friendly'' closed-form real energies coincided with their
elementary square-well $q= 2$ predecessors. In contrast,  it
appeared rather difficult to localize the precise position of all of
the {\em complex} bound-state energies in complex plane. A
sophisticated numerical approach appeared necessary for the purpose.

In the context of physics the potential phenomenological
applicability of our present family of toy-model quantum graphs may
be perceived as guided by the parallels with the $q=2$ special-case
square-well which represents one of the simplest available ${\cal
PT}-$symmetric models. This parallelism may be expected to include,
e.g., a potential relation between the present $q=2$ non-constant
level $E_0=E_0(\alpha)$ of Eq.~(\ref{spectrum}) and the similar
anomalous levels which are known to emerge in supersymmetric models
\cite{Cooper}. In the future, other possible parallels might also
appear reflecting, say, the preservation of a certain
complex-rotational symmetry of our present wave functions (cf.
Eq.~(\ref{brody})) or the related graph-inspired
permutation-transformation generalization of the concept of the
parity, etc.

In the context of mathematics, one of the most unexpected byproducts
of the transition to $q>2$ occurred at the subsequence of models
with $q=4m-2$. In contrast to the presence of a {\em single}
anomalous real energy level with $k = \alpha$ which existed at
$m=1$, it has been found that {\em infinitely many} anomalous real
energy levels seem to exist at any larger $m\geq 2$. This phenomenon
is a truly puzzling new structural feature of the spectrum of a
phenomenological model. Its deeper theoretical explanation (say, via
its possible relation to the complex-rotational symmetries of wave
functions) remains an open question at present.

\subsection*{Acknowledgements}

The support by the GA\v{C}R grant Nr. P203/11/1433 is acknowledged.


\newpage

\begin{thebibliography}{00}


\bibitem{BM}
E. Caliceti, S. Graffi and M. Maioli, Commun. Math. Phys. 75 (1980)
51;

V. Buslaev and V. Grecchi, J. Phys. A: Math. Gen. 26 (1993) 5541;

C. M. Bender and K. A. Milton, Phys. Rev. D 55 (1997) R3255.

\bibitem{Carl}
C. M. Bender, Rep. Prog. Phys. 70 (2007) 947.

\bibitem{sm}
A. Mostafazadeh,
Ann. Phys. (N.Y.) 309 (2004) 1;

M. Znojil,
J. Phys. A: Math. Gen. 37 (2004) 9557;

A. Mostafazadeh and F. Zamani,
%
Ann. Phys. (N.Y.) 321 (2006) 2183;

V. Jakubsk\'{y}, J. Smejkal,
Czech. J. Phys. 56 (2006) 985.

\bibitem{ali}
A. Mostafazadeh, Int. J. Geom. Meth. Mod. Phys. 7 (2010) 1191.

\bibitem{ptsusy}
%
A. Andrianov, M. V. Ioffe, F. Cannata and J.-P. Dedonder, Int. J.
Mod. Phys. A 14 (1999) 2675;

M. Znojil,
J. Phys. A: Math. Gen. 35 (2002) 2341;

%
%
B. Bagchi, S. Mallik and C. Quesne, Mod. Phys. Lett. A 17 (2002)
1651.

\bibitem{pts}
F. Correa, V. Jakubsk\'{y}, L. M. Nieto and M. S. Plyushchay,
Phys. Rev. Lett. 101 (2008) 030403;

F. Correa, V. Jakubsky, M. S. Plyushchay,  Annals Phys. 324 (2009)
 1078;

F. Correa and  M. S. Plyushchay,
 Annals Phys. 327 (2012) 1761.

\bibitem{Makris}
Z. H. Musslimani et al, Phys. Rev. Lett. 100 (2008) 030402.

\bibitem{david}
D. Krej\v{c}i\v{r}\'{\i}k, H. B\'{\i}la and M, Znojil,
%
J. Phys. A: Math. Gen. 39 (2006) 10143.
%

\bibitem{SIGMA}
M. Znojil, SIGMA 5 (2009) 001, eprint arXiv:0901.0700.

\bibitem{Siegl}
D.  Krej\v{c}i\v{r}\'{\i}k,
J. Phys. A: Math. Gen. 41 (2008) 244012;

D. Krej\v{c}i\v{r}\'{\i}k, P. Siegl and J. \v{Z}elezn\'y,
On the similarity of Sturm-Liouville operators with non-Hermitian
boundary conditions to self-adjoint and normal operators, submitted,
arXiv:1108.4946.

\bibitem{siegldis}
P. Siegl,
 Non-Hermitian quantum models, indecomposable
representations and coherent states quantization (PhD thesis, Univ.
Paris Diderot  \& FNSPE CTU, 2011);

J. \v{Z}elezn\'y, The Krein-space theory for non-Hermitian
PT-symmetric operators (MSc thesis, FNSPE CTU, 2011).

\bibitem{Sieglbe}
D. Krej\v{c}i\v{r}\'{\i}k and P. Siegl,
J. Phys. A: Math. Theor. 43 (2010) 485204;
%

D. Borisov and D. Krej\v{c}i\v{r}\'{\i}k,
Asympt. Anal. 76 (2012) 49;

D. Kochan, D.  Krej\v{c}i\v{r}\'{\i}k, R. Nov\'{a}k and P. Siegl,
J. Phys. A: Math. Theor., to appear, arXiv: 1203.5011


\bibitem{[4]} see, e.g., http://en.wikipedia.org/wiki/Quantum graph

\bibitem{[7]}
P. Kuchment, Waves in Random Media 14 (2004) S107.

\bibitem{[6]}
P. Exner, J. P. Keating, P. Kuchment, and A. Teplyaev (editors),
Analysis on Graphs and Its Applications (AMS, Rhode Island, 2008).

\bibitem{[12]}
T. Kottos, and U. Smilansky, Phys. Rev. Lett. 79, 4794 (1997);

S. Gnutzmann and U. Smilansky, Advances in Physics 55 (2006) 527.

\bibitem{[8]}
P. Exner, Ref. \cite{[6]}, p. 523.

\bibitem{Wigner}
E. Wigner, J. Math. Phys. 1 (1960) 409 and 414;

J. Dieudonne, Proc. Int. Symp. Linear Spaces, p. 115 (1961).

\bibitem{wax}
A. G. Anderson and C. M. Bender, Complex Trajectories in a Classical
Periodic Potential. Preprint arXiv:1205.3330;

C. M. Bender, B. K. Berntson, D. Parker and E. Samuel, Observation
of PT phase transition in a simple mechanical system. Preprint
arXiv:1206.4972.


\bibitem{unbou}
A. Mostafazadeh, Pseudo-Hermitian Quantum Mechanics with Unbounded
Metric Operators. Preprint arXiv:1203.6241; accepted for
publication; presented during the recent int. conference
``Non-Hermitian Operators in Quantum Physics'' (APC Paris, August 27
- 31, 2012, webpage http://phhqp11.in2p3.fr/Home.html).

\bibitem{Coron}
H. Hernandez-Coronado, D. Krej\v{c}i\v{r}\'{\i}k and P. Siegl,
Phys. Lett. A 375 (2011) 2149.


\bibitem{Krejcirik}
P. Siegl and D. Krej\v{c}i\v{r}\'{\i}k, Metric operator for the
imaginary cubic oscillator does not exist. Preprint arXiv:1208.1866;
presented during the recent int. conference ``Non-Hermitian
Operators in Quantum Physics'' (APC Paris, August 27 - 31, 2012,
webpage http://phhqp11.in2p3.fr/Home.html).

\bibitem{grII}
M. Znojil, 
Phys. Rev. D. 80 (2009) 105004.

\bibitem{int1}
M. Znojil,
%
J. Phys. A: Math. Theor. 43 (2010) 335303.

\bibitem{Calogero}
F. Calogero, J. Math. Phys. 10 (1969) 2191.

\bibitem{Klee}
V. Jakubsk\'{y},
Czech. J. Phys. 54 (2004) 67;

V. Jakubsk\'{y},  M. Znojil, E. A. Lu\'{\i}s and F. Kleefeld,
Phys. Lett. A 334 (2005) 154;

A. Fring and M. Znojil,
J. Phys. A: Math. Theor. 41 (2008) 194010;

F. Tremblay, A. V. Turbiner and P. Winternitz, 
J. Phys. A: Math. Theor. 42 (2009) 242001;

C. Quesne, J. Phys. A: Math. Theor. 43 (2010) 305202.

\bibitem{Cooper}
F. Cooper, A. Khare and U. Sukhatme, Phys. Rep. 251 (1995) 267.


\end{thebibliography}
\end{document}